\documentclass[runningheads,fleqn]{svmult}

\usepackage{makeidx}   
\usepackage{graphicx}  
\usepackage{subeqnar}  
\usepackage{multicol}  
\usepackage{physproc}  
\makeindex             

\begin{document}
\title*{Radiative spacetimes}
\toctitle{Radiative spacetimes}
%
%
\titlerunning{Radiative spacetimes}
%
\author{Ji\v r\' \i\ Bi\v c\' ak}
\authorrunning{Ji\v r\' \i\ Bi\v c\' ak}
%
%
\institute{Institute of Theoretical Physics,\\Charles University, Prague\\{\rm bicak@mbox.troja.mff.cuni.cz}}

\maketitle              

\begin{abstract}
The question of existence of general, asymptotically flat radiative spacetimes and examples of explicit classes of radiative solutions of Einstein's field equations
are discussed in the light of some new developments. The examples are
cylindrical waves,
Robinson-Trautman and type N spacetimes and especially boost-rotation symmetric spacetimes
representing uniformly accelerated particles or black holes.
\end{abstract}

\section{Introduction}
In physical theories on a fixed background spacetime, as in
Newtonian theory or special relativity, it is not difficult to
formulate asymptotic fall-off conditions on fields of spatially
bounded systems. For example, the gravitational
potential due to a Newtonian star is usually required to decay to
zero at infinity of Euclidean space, with the decay rate being
compatible with Laplace's equation. In general relativity no a
priori given
background space exists. The metric itself is both a dynamical
field and a quantity which determines distances. One expects that
in a suitable coordinate system far away from a system of bodies
the metric should have a form $g_{\mu \nu} = \eta_{\mu\nu} +$
small quantities, where $\eta_{\mu\nu}$ is Minkowski metric.
What, however, does it mean "far away", what is "infinity"? Can
one formulate suitable boundary conditions in a coordinate-free
manner? What is the decay of a {\it radiative} gravitational field?

After several important contributions to the gravitational
radiation theory in the late 1950's and early 1960's by Pirani,
Bondi, Robinson, Trautman and others, a landmark paper by Bondi
et al \cite {e} appeared in which radiative properties of isolated
(spatially bounded) axisymmetric systems were studied along
outgoing null hypersurfaces $u=constant$, with $u$ representing a
retarded time function. An ansatz was made that the metric along
$u=constant$ can be expanded in inverse powers of $r$,

\begin{equation}\label{RSE1}
g_{\mu v} = \eta_{\mu v}+h_{\mu v}(\theta) r^{-1}+f_{\mu
v}(\theta, \varphi)r^{-2}+\ldots,
\end{equation}
where $r$ denotes a suitable parameter along null generators
(parametrized by coordinates $\theta, \varphi$) on the
hypersurfaces $u=constant$. Under the assumption (\ref{RSE1}) Einstein's
vacuum equations were shown to determine uniquely formal power
series solution of the form (\ref{RSE1}), provided that a free "news
function" $c(u, \theta)$ is specified. The news function contains
all information about radiation at infinity $("r=\infty")$. It
enters the fundamental "Bondi mass-loss formula" for the total
mass $M(u)$ of an isolated system at retarded time $u$. Field
equations imply that $M(u)$ is a monotonically decreasing
function of $u$ if $\partial_u c \not= 0.$ A natural
interpretation is that gravitational waves carry away positive
energy from the system and thus decrease its mass. In the work of
Bondi et al \cite{e} as well as in the important generalizations by
Sachs, Newman and Penrose, the decay of radiative fields was studied in
preferred coordinate systems.

In 1963 Penrose \cite{h} formulated a beautiful {\it
geometrical} framework for description of the "radiation zone"
in general relativity in terms of conformal infinity.

Penrose's definition of asymptotically flat radiative
spacetimes avoids such problems as "distances" or "suitable
coordinates", and incorporates a clear definition of what is
infinity. It is inspired by the work on radiation theory
mentioned above, and by the properties of conformal infinity in Minkowski
spacetime. In contrast to an Euclidean space, in Minkowski
spacetime one can go to infinity in various directions: moving along
timelike geodesics we come to the future (or past) timelike
infinity $I^+$ (or $I^-$); along null geodesics (cf. Eq. (\ref{RSE1})) we
reach the future (past) null infinity ${\cal J}^+ ({\cal J}^-)$; spacelike
geodesics lead to spatial infinity ${\it i}_0$. Minkowski spacetime
can be compactified and mapped into a finite region by an
appropriate conformal transformation. Thus one obtains the
well-known {\it Penrose diagram} in which the three types
of infinities are mapped into the boundaries of the compactified
spacetime - see Fig. 1.

It is generally accepted that Penrose's definition forms the only
rigorous, geometrical basis for the discussion of gravitational
radiation from isolated systems. It enables us to use techniques
of local geometry "at infinity" and to define covariantly such
fundamental quantities as the total (Bondi) mass of an isolated
system.

\begin{figure}
\includegraphics[width=274pt]{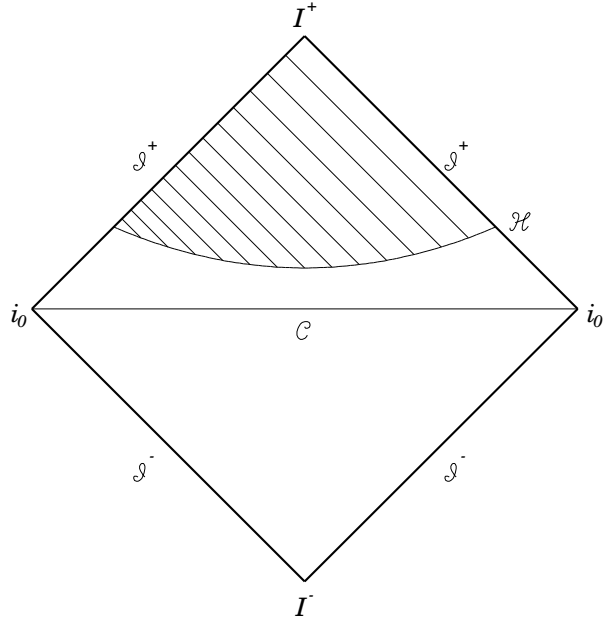}
\caption[]{The Penrose conformal diagram of an asymptotically flat spacetime. The Cauchy hypersurface and the hyperboloidal hypersurfaces are indicated.}
\label{Fig1}
\end{figure}

\section{Asymptotically flat radiative spacetimes: existence}

Despite its rigour and elegance, Penrose's {\it definition} might turn
to be of a limited importance if no interesting radiative
spacetimes {\it exist} which satisfy the definition. In Section 5
we shall describe special exact radiative spacetimes which
represent "uniformly accelerated" sources in general relativity
and admit $\cal J$ as required; however, at least four points on $\cal J$ -
those in which worldlines of the sources start and end - are
singular. There are no other {\it explicit} exact radiative solutions
describing finite sources available at present and this situation
will probably not change soon. Nevertheless, thanks to the work
of Friedrich \cite{HF3} and Christodoulou and Klainerman \cite{l} we know that
globally non-singular (including {\it all} $\cal J$ ) asymptotically flat
exact solutions of Einstein's equations really exist.

The key idea of Friedrich is in realizing that Penrose's
treatment of infinity not only permits to use the methods of
local differential geometry at $\cal J$ but also to analyze
global existence problems of solutions of Einstein's equations in
physical spacetime $M$ by solving initial value problems in
unphysical spacetime $\tilde M $.

By using this approach Friedrich established that formal
Bondi-type expansions (\ref{RSE1}) converge locally at ${\cal {J}}^+$.
He succeeded to show that
one can formulate the "hyperboloidal initial value problem" for
Einstein's vacuum equations in which initial data are given on a
hyperboloidal spacelike hypersurface $\cal H$ which intersects $\cal J$
(see Fig. 1). It can then be proven that hyperboloidal initial
data, which are sufficiently close to Minkowskian hyperboloidal
data (i.e. to the metric induced on the hypersurface $\cal H$ in
Minkowski spacetime by the standard Minkowski metric) evolve to
a vacuum spacetime
which is smooth on $\cal J^+$ and ${\it I}^+$ as required by
Penrose's definition.

Despite the deep and complicated work by Friedrich, one still did
not have what one really would wish. Let us first note that
initial data {\it "sufficiently close"} to Minkowskian data in
Friedrich's result we mentioned above, do not mean any
approximation - "smallness" of data is understood in a functional
sense in an appropriate Sobolev space. One cannot hope to prove
global existence of smooth general solutions developed from
general, arbitrarily
{\it strong} data. Vacuum data representing strong gravitational waves
could, due to nonlinearities, lead to creation of black holes
with singularities inside. Nevertheless, ultimately one would
like to have also results on the evolution of strong data.

More importantly, however, we would like to have initial data
given on a {\it standard} spacelike Cauchy hypersurface which
does not intersect $\cal J$ but "ends" at spatial infinity
(cf. hypersurface $\cal C$ in Fig. 1),
rather than data given on a
hyperboloidal initial hypersurface. It could well happen that
a spacetime evolved from
data on $\cal H$ is smooth "above" $\cal H$
(in the shaded region in Fig. 1), however, it does
not satisfy Penrose's requirements of asymptotic flatness
"bellow" $\cal H$.

A remarkable progress in proving rigorously the existence of
general, asymptotically flat radiative spacetimes was achieved by
Christodoulou and Klainerman. Their treatise \cite{l} contains
the first really global general existence statement for full,
nonlinear Einstein's vacuum equations with vanishing
cosmological constant: Any smooth asymptotically flat initial
data set (determined by the first and the second fundamental
form on a Cauchy hypersurface) which is "near flat (Minkowski)
data" leads to a unique, smooth and geodesic complete development
solution of Einstein's vacuum equations. This solution is
"globally asymptotically flat" in the sense that the curvature
tensor decays to zero at infinity in all directions. The
Christodoulou-Klainerman theorem involves a "global smallness
assumption"
which requires appropriate (integral) norms of curvature tensor
of initial data integrated over the initial Cauchy hypersurface
to be small.

The theorem demonstrates the existence of singularity-free,
asymptotically flat
radiative vacuum spacetimes. In hydrodynamics, even for arbitrary small
initial data (decaying at infinity), an analogous theorem is not
true since shocks arise. In general relativity the effect of
nonlinear terms, which could have led to formation of
singularities, is excluded owing, apparently, to the covariance
and algebraic properties of the Einstein vacuum equations. That
singularities could have had well developed is evident:
the collision of arbitrarily weak,
vacuum plane gravitational waves leads (due to nonlinearities)
to the formation of a singularity. This, of course, does not
contradict the Christodoulou-Klainerman result because initial
data for plane waves are not asymptotically flat.

The work of Friedrich, Christodoulou, Klainerman and others
demonstrates
rigorously that the general picture of null infinity is
compatible with the vacuum Einstein field equations. However,
important open questions remain.  In classical
papers by Bondi et al \cite{e}, Newman and Penrose
and others, the decay of
the curvature (characterized by
the Weyl tensor) along outgoing null
geodesics at infinity exhibits "the peeling-off" property: the
fall-off of various components of the Weyl tensor is related to
their Petrov algebraic type. To be more specific, certain complex
linear combinations of the Weyl tensor in the orthonormal frame,
$\Psi_k (k=0,1,2,3,4)$, behave as $\Psi_k = O (r^{k-5})$ as $r
\rightarrow \infty$. (In particular, $\Psi_4 \sim r^{-1}$ has the
same algebraic structure as the Weyl tensor of a plane wave --
radiative field of a bounded system resembles asymptotically
that of a plane wave.)  This decay of the curvature can be shown to follow
from a sufficient differentiability
(smoothness) of the conformally rescaled
(unphysical) metric $\tilde g$. A sufficient smoothness of null
infinity is thus commonly assumed. The results
of Christodoulou and Klainerman, however, show a weaker peeling.
They were only able to prove that the asymptotically flat vacuum
initial data lead to $\Psi_0 \sim r^{-\frac{7}{2}}$ ({\it not}
$\sim r^{-5}$) at null infinity.

An increasing evidence against the proposal of a smooth $\cal J$
has led Chru\'{s}ciel et al \cite{o} to introduce the
concept of a {\it polyhomogeneous} $\cal J$. The metric is
called polyhomogeneous if at large {\it r} it admits an expansion in
terms of $r^{-j}\log^i r$ rather than $r^{-j}$ (as it has been
assumed in the works of Bondi and others - cf. Eq. (\ref{RSE1})). The
hypothesis of polyhomogeneity of $\cal J$ has been shown to be
formally consistent with Einstein's vacuum equations. Under
appropriate assumptions on the asymptotic form of the
polyhomogeneous metric, one can demonstrate that the Bondi
mass-loss law can be formulated, and the peeling-off property of
the curvature holds, with the first two terms identical to the
standard peeling, the third term being $\sim r^{-3} \log r$. At
null infinity the conformally rescaled (unphysical) metric is
not smooth. Although a substantial
progress in understanding the existence of radiative solutions of
vacuum field equations and the asymptotic structure of
corresponding radiative spacetimes has been achieved, we have
seen that
open problems remain. Curiously enough, in the case of vacuum
Einstein's
equations with a {\it non-vanishing cosmological constant}
a more complete picture is known for some time already. By using his regular conformal field equations, Friedrich
demonstrated \cite{y} that initial data sufficiently close to
de-Sitter data develop into solutions of Einstein's equations
with a positive cosmological constant, which are asymptotically
simple (with a smooth conformal infinity), as required in the
original framework of Penrose. Later
Friedrich \cite{z} also discussed the existence of asymptotically simple
solutions to the Einstein vacuum equations with a negative
cosmological constant.

In his more recent investigations \cite{HF3}, Friedrich constructed
the new - finite but "wider" than the point $i_0$ - representation of
spacelike infinity. This construction enables one to make much
deeper analysis of the initial data in the region where null
infinity touches spacelike infinity. Good chances now exist to
obtain clear criteria determining which data lead to the smooth
and which just to the polyhomogeneous null infinity.

The ultimate goal of rigorous work on the existence and asymptotics
of solutions of the Einstein equations is {\it physics}: one
hopes to be able to consider astrophysical sources, to relate
their behaviour to the characteristics of the far fields. One
would like to have under control various (both analytical and
numerical) approximation procedures. A still more ambitious
program
is to consider strong initial data so as to be able to analyze
such issues as cosmic censorship.

In the following we shall briefly discuss three classes of
{\it explicit} radiative solutions of Einstein's equations: cylindrical
waves, Robinson-Trautman and type N spacetimes, and the
boost-rotation symmetric spacetimes. The latter represent the only
known examples describing moving, radiating objects; except for
points of null infinity where particles start and end, the null
infinity is smooth. We here closely follow our recent, more
detailed reviews \cite{BISP,BIJOUR} in which also other classes of radiative
spacetimes are analyzed, such as plane waves and gravitational
waves representing inhomogeneous cosmological models.

\section{Cylindrical waves}

Despite the fact that cylindrically symmetric waves cannot
describe exactly the radiation from bounded sources, they even recently played
an important role in clarifying a number of complicated issues,
such as testing the quasilocal mass-energy, testing
codes in numerical relativity, investigation of the
cosmic censorship, and quantum gravity.

In work with Ashtekar and Schmidt \cite{ABS1,ABS2},
we considered gravitational waves with a
space-translation Killing field (``generalized Einstein-Rosen
waves''). In (2+1)-dimensional framework the
Einstein-Rosen subclass forms a simple instructive example of
explicitly given spacetimes which
admit a smooth global null (and timelike) infinity even for
strong initial data.

4-dimensional vacuum gravity which admits a spacelike hypersurface Killing vector
$\partial / {\partial z}$ is equivalent to 3-dimensional
gravity coupled to a scalar field. In 3 dimensions, there is no
gravitational radiation. Hence, the local degrees of freedom are
all contained in the scalar field. One therefore expects that
Cauchy data for the scalar field will suffice to determine the
solution. For data which fall off appropriately, we thus expect
the 3-dimensional Lorentzian geometry to be asymptotically flat
in the sense of Penrose, i.e. that there
should exist a 2-dimensional boundary representing null infinity.
In general cases, this is analyzed in \cite{ABS1}.

Restricting ourselves to the Einstein-Rosen waves by assuming that
there is a further spacelike, hypersurface orthogonal Killing
vector $\partial /\partial\varphi$ which commutes with
$\partial/\partial z$, we find the 3-metric given by
\begin{equation}
\label{Equ70}
d\sigma^2= g_{ab}dx^adx^b = e^{2\gamma}(-dt^2+d\rho^2)+
\rho^2 d\varphi^2 .
\end{equation}
The field equations become
\begin{equation}
- \ddot\psi + \psi'' + \rho^{-1}\psi' = 0~,~~~
\gamma' = \rho (\dot{\psi}^2+\psi'^2)~, ~~~
\dot \gamma = 2\rho \dot \psi \psi'.
\end{equation}
Thus, we can first solve the axisymmetric wave equation
for $\psi$ on Minkowski space and then solve for
$\gamma$ -- the only unknown metric coefficient -- by quadratures.

By analyzing the asymptotic behavior of the solutions we can conclude that {\it cylindrical waves in
(2+1)-dimensions give an explicit model of the
Bondi-Penrose radiation theory which admits smooth null and timelike
infinity for arbitrarily strong initial data}.
There is no other such model available.
The general results on the existence of $\cal J$ in 4
dimensions assume weak data.

\section{On the Robinson-Trautman and type N twisting solutions}

These spacetimes have attracted increased attention in the
last decade -- most notably in the work by Chru\'{s}ciel, and
Chru\'{s}ciel and Singleton \cite{ak}. In these studies the
Robinson-Trautman spacetimes have been shown to exist globally for all
positive ``times'', and to converge asymptotically to a
Schwarzschild metric. Interestingly, the extension of these
spacetimes across the ``Schwarz\-schild-like'' event horizon can only
be made with a finite degree of smoothness. These
studies are based on the derivation and analysis of an asymptotic
expansion describing the long-time behaviour of the solutions of
the nonlinear parabolic Robinson-Trautman equation.

In our work \cite{BiPo,al} we studied Robinson-Trautman
spacetimes with a positive cosmological constant $\Lambda$. The
results proving the global existence and convergence of the
solutions of the Robinson-Trautman equation can be taken over
from the previous studies since $\Lambda$ does not explicitly
enter this equation. We have shown that,
starting with arbitrary, smooth initial data at $u=u_0$,
these cosmological Robinson-Trautman solutions converge exponentially fast
to a Schwarzschild-de Sitter solution at large retarded times
($u\to \infty$).
The interior of a
Schwarzschild-de Sitter black hole can be joined to an ``external''
cosmological Robinson-Trautman spacetime across the horizon
$\cal H^+$ with
a higher degree of smoothness than in the corresponding case with
$\Lambda = 0$. In particular, in the extreme case with
$9 \Lambda m^2 = 1$, in which the black hole and cosmological horizons coincide,
the Robinson-Trautman spacetimes can be extended smoothly
through $\cal H^+$ to the extreme Schwarzschild-de Sitter spacetime
with the same values of $\Lambda$ and $m$. However, such an extension
is not analytic (and not unique).

We have also demonstrated that the cosmological
Robinson-Trautman solutions represent explicit models exhibiting
the cosmic no-hair conjecture. As far as we are aware,
these models represent the only exact analytic demonstration of the cosmic
no-hair conjecture under the presence of gravitational waves.
They also appear to be the only exact examples of black hole
formation in nonspherical spacetimes which are not asymptotically
flat.

\subsection*{Type N twisting spacetimes}

Since diverging, non-twisting Robinson-Trautman
spacetimes of type N have singularities, there has been hope
that if one admits a nonvanishing twist a more realistic
radiative spacetime may exist.

Stephani \cite{STP}, however, indicated, by constructing a general
solution of the linearized equations, that singularities at infinity
probably exist. More recently, Finley et al \cite{FIL} found
an approximative twisting type N solution up to the third
order of iteration on the basis of which they suggested that it
seems that the twisting, type N fields can describe a
radiation field outside bounded sources. However, employing
the Newman-Penrose formalism and MAPLE we succeeded in discovering a
nonvanishing quartic invariant in the 2nd derivatives of the
Riemann tensor \cite{BiPr},
which shows that solutions of both Stephani and Finley et al
contain singularities at large $r$. Mac Alevey
\cite{MA} argued that an approximate solution at any finite order
can be calculated without occurrence of singularities. It is very
likely, however, that a corresponding exact solution must contain
singularities since Mason \cite{MASO} proved that the
only vacuum algebraically special spacetime that is
asymptotically simple is the Minkowski space.

Even if a radiative solution with complete smooth null infinity
may be out of reach, it is of interest to construct
radiative solutions which at least admit a global null infinity in
the sense that its smooth cross sections exist although this null
infinity is not necessarily complete. The only explicit examples
of such solutions are spacetimes with boost-rotation symmetry.

\begin{figure}
\includegraphics[width=365pt]{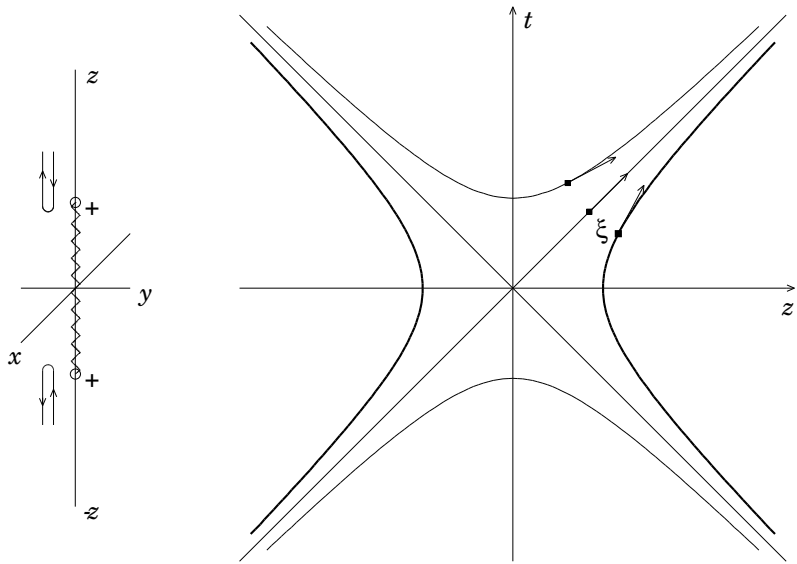}
\caption[]{Two particles uniformly accelerated in opposite directions.}
\label{Fig2}
\end{figure}

\begin{figure}
\centering
\includegraphics[width=.71\textwidth]{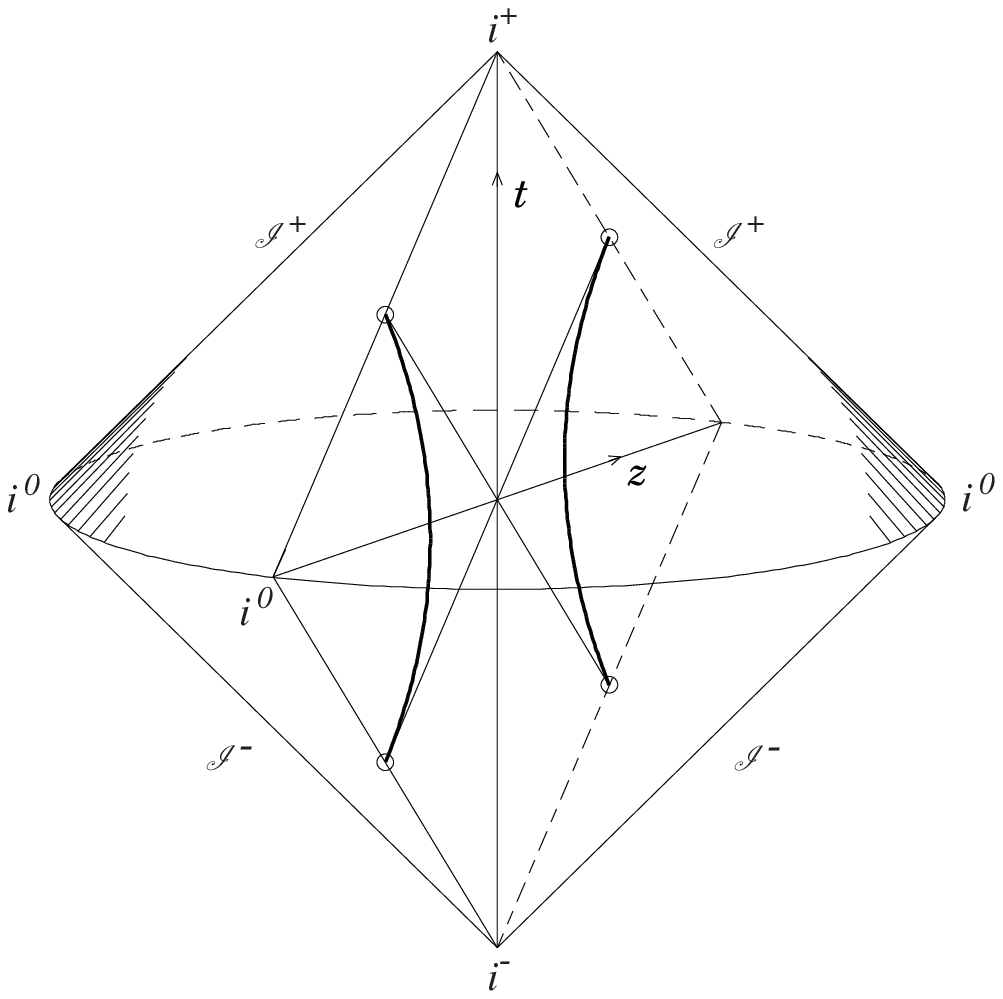}
\label{Fig3}
\caption{The Penrose compactified diagram of a boost-rotation symmetric spacetime. Null infinity can admit smooth sections.}
\end{figure}

\section{The boost-rotation symmetric radiative spacetimes}

I reviewed these spacetimes representing ``uniformly accelerated objects'' (see Fig. 2) in various places
(see e.g. \cite{BISP,KVB} and references therein);
here I shall just mention some new results. The Penrose diagram of
these spacetimes is schematically illustrated in Fig. 3.

The unique role of the boost-rotation symmetric spacetimes is exhibited by a
theorem \cite{BPa} which roughly states that  in {axially} symmetric, locally
asymptotically flat electrovacuum spacetimes (in the sense that a null infinity
satisfying Penrose's requirements exists, but it need not
necessarily exist globally), the only {additional} symmetry
that does not exclude radiation is the {\it boost} symmetry.

To prove such result we start from the metric

\begin{eqnarray}
\label{ds}
ds^2&=&{\left(r^{-1}\, V\, {{e}^{2\beta}}\!-\!r^2{e^{2\gamma }} U^2\,{\cosh 2\delta}
           \!-\!r^2{e^{-2\gamma}} W^2\,{\cosh 2\delta}\!-\!2r^2 UW\,{\sinh 2\delta}\right)}\ du^2\nonumber\\
    & &\ + 2{e^{2\beta}} du~dr
              +2r^2\left({e^{2\gamma }} U\,{\cosh 2\delta} +W\,{\sinh 2\delta}\right)\ du\ d\theta\nonumber\\
    & &
        \ +2r^2\left({e^{-2\gamma}} W\,{\cosh 2\delta} +U\,{\sinh 2\delta}\right){\sin \theta}\ du\ d\phi\\
    & &\ - r^2\left[ {\cosh 2\delta}\left({e^{2\gamma }} { d}\theta^2 +{e^{-2\gamma}} {\sin^2 \theta}\ d\phi^2 \right)
                        +2\,{\sinh 2\delta}\ {\sin \theta}\ d\theta\ d\phi\right] \ ,\nonumber
\end{eqnarray}
where all funtions describing metric and electromagnetic field tensor $F_{\mu \nu}$ are independent of $\phi$.
Assuming asymptotic expansions of these functions at large $r$ with $u$, $\theta$, $\phi$ fixed
to guarantee asymptotic flatness, and using the outgoing radiation condition and the field equations,
one finds these expansions to have specific forms. For example,

$$
\gamma  =\frac{c}{r}+( C-{{\scriptstyle{\frac{1}{6}}}}c^3
             -{{{\scriptstyle{\frac{3}{2}}}}} cd^2)
           \frac{1}{r^3}+...\ ,\
~~~~~~V   = r - 2M  + ...~,
$$
\begin{equation}
~~~~~~~~~F_{02} = X\;+\;(\epsilon_{,\theta}-e_{,u}){1\over r} + ...\ ,~~~~~~~~~~~~~F_{03} = Y\;-\;{f_{,u}\over r} + ...~,
\label{jednarovnice}
\end{equation}
where the `coefficients' $c$, $d$, ... are functions of $u$ and $\theta$. The expansions are needed
to further orders -- see \cite{BPa} for their forms. Let us only recall that the decrease
of the Bondi mass,
$
m(u)=
{1 \over 2}\int_{0}^{\pi} M(u,\theta)\sin \theta d\theta\ ,\label{hmota}
$
is given by
\begin{equation}
m_{,u} =-{{{\scriptstyle{\frac{1}{2}}}}}\int\limits_{0}^{\pi} (c,_u^2+d,_u^2+X^2+Y^2)\sin \theta d\theta
               \leq 0\ ,  \label{klhmota}
\end{equation}
where $c_{,u}$, $d_{,u}$, $X$, $Y$ are gravitational and electromagnetic news functions.

Now one writes down the Killing equations and solves them asymptotically in $r^{-1}$. One arrives
at the following theorem \cite{BPa}: Suppose that an axially symmetric
electrovacuum spacetime admits a ``piece'' of ${\cal J}^+$ in the~sense
that the~Bondi-Sachs coordinates can be introduced in which
the~metric takes the~form (\ref{ds}), with
the~asymptotic form of the~metric and electromagnetic field given by
(\ref{jednarovnice}). If this spacetime admits an additional Killing
vector forming with the~axial Killing vector a two-dimensional Lie
algebra then the~additional Killing vector has asymptotically
the~form
\begin{equation}
\eta^\alpha=[-ku\cos \theta+\alpha(\theta),\
          kr\cos \theta+{\cal O}(r^{0}),\ -k\sin \theta+{\cal O}(r^{-1}),\ {\cal O}(r^{-1})]\ ,
\label{Bbotr}
\end{equation}
where $k$ is a constant. For $k=0$ it generates asymptotically translations
(function $\alpha$ has then a specific form). For $k\not= 0$ it is the~boost Killing field.

The case of translations is analyzed in detail in \cite{BAP}.
Theorem 1, precisely formulated and proved there, states that if
asymptotically translational Killing vector is spacelike, then
null infinity is singular at some $\theta \not= 0, \pi$; if it is
null, null infinity is singular at $\theta = 0$ or $\pi$. The
first case corresponds to cylindrical waves, the second case to a
plane wave propagating along the symmetry axis. We refer to
\cite{BAP} for the case when there is also a cosmic string
present along the symmetry axis. The case of timelike Killing
vector is described by Theorem 2 (proved also in \cite{BAP}):
``If an axisymmetric electrovacuum spacetime with a non-vanishing
Bondi mass admits an asymptotically translational Killing vector
and a complete cross section of ${\cal J}^+$, then the translational
Killing vector is timelike and spacetime is thus stationary.''

The case of the boost Killing vector $(k\not=0)$ is thoroughly
analyzed in \cite{BPa}. The general functional forms of the news
functions (both gravitational and electromagnetic), and of the
mass aspect and total Bondi mass of boost-rotation symmetric
spacetimes are there given. Recently these results were obtained \cite{VKR} by using the
Newman-Penrose formalism and under more general assumptions
(for example, {$\cal J$} could in principle be polyhomogeneous).

The general structure of the boost-rotation symmetric spacetimes
with hypersurface orthogonal Killing vector was analyzed in
detail in \cite{ao}. Their radiative properties, including
explicit  construction of radiation patterns and of Bondi mass
for the specific boost-rotation symmetric solutions were
investigated in several works --
we refer to the reviews \cite{BISP,KVB} and \cite{PPRD}
for details. There also the role of the boost-rotation symmetric
spacetimes in such diverse fields like numerical relativity and
quantum production of black-hole pairs is noticed and
references are given.

Here I would like to mention yet a recent progress in
understanding specific boost-rotation symmetric spacetimes with
Killing vectors which are {\it not} hypersurface
orthogonal. This is the {\it spinning} $C$-metric (see e.g. \cite{KSH}). It was
discovered by Pleba\'nski and Demai\'nski as a generalization of the
standard $C$-metric which is known to represent uniformly
accelerated non-rotating black holes. In \cite{BPD} we first
transformed the metric into Weyl coordinates, and then found
a transformation which brings it into the canonical form
of the radiative spacetimes with the boost-rotation symmetry:

\begin{eqnarray}
\label{BStvarR}
{ ds}^2 &=& { e}^{\lambda} { d} \rho^2 + \rho^2 { e}^{-\mu} { d} \phi^2 \nonumber  \\
 & + & {(z^2-t^2)^{-1}} \left[ ({ e}^{\lambda} z^2 - { e}^{\mu} t^2 ) { d} z^2
  -     2zt ({ e}^{\lambda} - { e}^{\mu}  ) { d} z\  { d} t \right.\nonumber\\
   & + & \left.    ({ e}^{\lambda} t^2 - { e}^{\mu} z^2 ) { d} t^2  \right]
~  -     2{\cal A} { e}^{\mu} (z { d}t -  t { d} z)  { d} \phi -{\cal A}^2 { e}^{\mu} (z^2-t^2)   { d} \phi^2 \ ,
\end{eqnarray}
where functions $e^{\mu}, e^{\lambda}$ and ${\cal A}$ are given in terms of $(t,
\rho, z)$ in a somewhat complicated but explicit manner.
This metric can represent two uniformly
accelerated, spinning black holes, either connected by a conical
singularity, or with conical singularities extending from each of
them to infinity. The behaviour of curvature
invariants clearly indicates the presence of a non-vanishing
radiation field (see Figure 5 in \cite{BPD}). The spinning
$C$-metric is the only explicitly known example
with two Killing vectors which are not hypersurface orthogonal,
in which one can give arbitrarily strong initial data on hyperboloid ``above
the roof" ($t>|z|$) which evolve into the radiative spacetime
with smooth ${\cal J}^+$.

%

\end{document}